\RequirePackage{fix-cm}
\documentclass[twocolumn]{svjour3}          
\smartqed  
\usepackage{graphicx}
\usepackage{amsmath}
\begin{document}

\title{Measurement and Collapse within the Two-State-Vector Formalism
}


\author{Yakir Aharonov         \and
        Eliahu Cohen \and
        Eyal Gruss \and
        Tomer Landsberger
}


\institute{Y. Aharonov, E. Cohen, T. Landsberger \at
              School of Physics and Astronomy, Tel Aviv University, Tel Aviv 6997801, Israel. \\
              Tel.: +972-3-6408334\\
              Fax: +972-3-6407932\\
              \email{yakir@post.tau.ac.il, eliahuco@post.tau.ac.il, tomerlan@post.tau.ac.il}           
             \emph{Second address of Y. Aharonov: Schmid College of Science, Chapman University, Orange, CA 92866, USA.} 
           \and
           E. Gruss \at
              eyalgruss@gmail.com
}

\date{Received: 21.04.2014 / Accepted: date}

\maketitle

\begin{abstract}
The notion of collapse is discussed and refined within the Two-State-Vector Formalism (TSVF). We show how a definite result of a measurement can be fully determined when considering specific forward and backward-evolving quantum states. Moreover, we show how macroscopic time-reversibility is attained, at the level of a single branch of the wavefunction, when several conditions regarding the final state and dynamics are met, a property for which we coin the term ``classical robustness under time-reversal''. These entail a renewed perspective on the measurement problem, the Born rule and the many-worlds interpretation.

\end{abstract}

\section{Introduction}
\label{intro}

Alongside its enormous empirical success, the quantum mechanical account of physical systems imposes a myriad of divergences from our thoroughly ingrained classical ways of thinking. These divergences, while striking, would have been acceptable if only a continuous transition to the classical domain were at hand. Strangely, this is not the case. The difficulties involved in reconciling the quantum with the classical have given rise to different interpretations, each with its own shortcomings. Traditionally, the two domains are sewed together by invoking an {\it ad hoc} theory of measurement, which has been incorporated in the axiomatic foundations of quantum theory. The ``Copenhagen school'' postulates a fundamental dualism between the microscopic quantum realm and the macroscopic classical realm, which leads to an effective theory dividing the world into quantum systems and classical measuring devices, but not defining a condition for belonging to one domain or the other.\\

The vague border between the classical and quantum realms gives rise to the well-known measurement problem. The problem is best understood by considering the unique properties of the quantum state space, which is boosted in size compared to the classical phase space, in order to accommodate distinctly non-classical entangled states and states of superposition. The former entails nonlocal correlations; the latter, stemming from the mutual incompatibility of conjugate observables, implies that the quantum reality cannot be accounted for in classical terms of definite physical properties. \\

However, a superposition is never observed directly - a measurement will yield one definite value of a physical property even when the state in not an eigenstate of the measured observable (recall Schr\"{o}dinger's famous cat, which is always found to be either dead or alive, but not both). In such cases, nothing in the quantum description dictates the exact result of a measurement. Textbook QM supplements the unitary evolution of Schr\"{o}dinger equation (SE) with a second dynamical law, which spells a non-unitary
break in the evolution upon measurement, a.k.a. collapse, physically changing the state of the system, instantaneously, to an eigenstate of the measured observable. Accordingly, the result will only be determined probabilistically, where the probability is given by the square amplitude of the eigenstate term, a postulate known as the Born rule. This is in stark contrast to classical mechanics, which only exhibits probabilities stemming from ignorance about the exact phase space state of the system, while remaining fully deterministic and local at the fundamental level of the physical laws.\\

While the collapse postulate makes QM perfectly operational, it introduces ambiguity into the theory. Given that any macroscopic object is just an aggregate of microscopic objects, as suggested by the lack of criterion for otherwise distinguishing them, it is not clear why the SE should not suffice for the full dynamical description of any process in nature. And if it were all encompassing, QM would have been deterministic, and not probabilistic, as the collapse and the Born rule maintain. Attempts to give satisfactory explanations to this predicament lead to discussions about the completeness of the quantum description, and different interpretations. The different approaches range from collapse theories such as the \hyphenation{}Ghirardi-Rimini-Weber (GRW) Spontaneous Localization Model  \cite{GRW} and successors thereof, through deterministic variable theories such as Bohmian mechanics  \cite{Bohm}, the objective general-relativistic collapse suggested by Penrose \cite{Penrose}, and all the way to the relative state interpretation by Everett \cite{EMWI}, which assumes nothing other than the standard axioms. In the latter, the different branches of a superposition are said to represent different co-existing states of reality, where the observation of a certain outcome is attributed to the specific state of the observer that is correlated to it in the superposition. Each of the superposition terms constitutes a ``branching world'', and is part of a Universal deterministically evolving wavefunction. Hence, it is also known as the ``Many Worlds Interpretation'' \cite{VMWI}.\\

The novel approach we shall present tries to tackle the difficulties without resorting to the usual notion of collapse. It will inherit the advantages of the MWI without assuming multiple realities. This approach suggests that a complete description of the physical state has to include two state-vectors, forming the ``two-state'' or the more general ``two-state density matrix'' (which will be described in detail in Sec. \ref{sec:2}). The states evolve independently by the same unitary dynamical law (and same Hamiltonian), but in opposite temporal directions (where the forward direction is defined according to the direction of entropy increase in the observed Universe). Whenever macroscopic objects, i.e. many-particle systems, are entangled with a microscopic system, as in a measurement, the setup/environment selects a preferred basis, while the unknown backward-evolving state selects a definite outcome from the known forward-evolving state, giving rise to a single definite physical reality. Thus, the probabilistic nature of quantum events can be thought of as stemming from our ignorance of the backward-evolving state, reintroducing the classical concept of probability as a measure of knowledge.\\

The decoherence program \cite{Zurek,Deco} has been successful in reducing, locally, the unobserved coherent superposition of macrostates into a mixture of effectively classical states, pointer states. The damping of the interference terms in the pointer states basis is attributed to the near orthogonally of environmental states entangled with them. By tracing out the environmental degrees of freedom, one may unveil the mixed state in which the system and apparatus are given. However, the trace operation is a purely mathematical procedure, which indicates no reduction of the global state to a single definite measurement outcome. Using the backward-evolving state of the TSVF, we shall demonstrate how a selection of a single outcome may be achieved.\\

We will focus our attention on the boundary conditions posed in each realm. In classical mechanics, initial conditions of position and velocity for every particle fully determine the time evolution of the system. Therefore, trying to impose a final condition would either lead to redundancy or inconsistency with the initial conditions. This situation is markedly different in the realm of quantum mechanics. Because of the uncertainty principle, an initial state-vector does not determine, in general, the outcome of a future measurement. However, adding another constraint, namely, the final (backward-evolving) state-vector, results in a more complete description of the quantum system in between these two boundary conditions, that has bearings on the determination of measurement outcomes. The usefulness of the backward-evolving state-vector was demonstrated in the works of Aharonov {\it et al.} \cite{ABL,AAV86,Reznik,Born,TSVF,Gruss1,Gruss2,AR,Quanta,unified,future,unusual,past,each}.\\

The emergence of specific macrostates seems non-unitary from a local perspective, and constitutes an effective ``collapse'', a term which will be used here to denote macroscopic amplification of microscopic events, complemented by a reduction via the final state. We will show that a specific final state can be assigned so as to enable macroscopic time-reversal or ``classical robustness under time-reversal'', that is, reconstruction of macroscopic events in a single branch, even though ``collapses'' have occurred. An essential ingredient in understanding the quantum-to-classical transition is the robustness of the macrostates comprising the measuring apparatus, which serves to amplify the microstate of the measured system and communicate it to the observer. The robustness guarantees that the result of the measurement is insensitive to further interactions with the environment. Indeed, microscopic time-reversal within a single branch is an impossible task because evolution was not unitary. Macroscopic time-reversal, which is the one related to our every-day experience, is possible, although non-trivial.\\

The rest of the paper is organized as follows: In Sec. \ref{sec:2} the TSVF is briefly described and our measurement model is motivated. The heart of the paper resides in Sec. \ref{sec:3}, where we scrutinize the subject of measurement and ``classical robustness'' through the prism of TSVF. Sec. \ref{Disc} concludes the paper.

\section{The Two-States-Vector Formalism}
\label{sec:2}
The basis for a time-symmetric formulation of quantum mechanics was laid in 1964 by Aharonov, Bergman and Lebowitz (ABL) who derived a probability rule concerning measurements performed on systems, with a final state specified in addition to the usual initial state \cite{ABL}. Such a final state may arise due to a post-selection, that is, performing an additional measurement on the system and considering only the cases with the desired outcome. Alternatively, some systems in nature may have an inherent final boundary condition, just as all systems have an initial boundary condition. Given an initial state $\Psi_{i}$ and a final state $\Psi_{f}$ , the probability that an intermediate measurement of the non-degenerate operator $A$ yields an eigenvalue $a_{k}$ is
\begin{equation} \label{ABL}
\begin{array} {lcl}
Pr\left(a_{k}|\Psi_{i},\Psi_{f}\right)=\frac{Pr\left(\Psi_{f}|a_{k}\right)Pr\left(a_{k}|\Psi_{i}\right)}{\sum_{j}Pr\left(\Psi_{f}|a_{j}\right)Pr\left(a_{j}|\Psi_{i}\right)}\\ \\
=\frac{\left|\left\langle \Psi_{f}|a_{k}\right\rangle \right|^{2}\left|\left\langle a_{k}|\Psi_{i}\right\rangle \right|^{2}}{\sum_{j}\left|\left\langle \Psi_{f}|a_{j}\right\rangle \right|^{2}\left|\left\langle a_{j}|\Psi_{i}\right\rangle \right|^{2}}
\end{array}
\end{equation}
For simplicity, no self-evolution of the states is considered between the measurements. If only an initial state is specified, Eq. \ref{ABL} should formally reduce to the regular probability rule:
\begin{equation} \label{rABL}
Pr\left(a_{k}|\Psi_{i}\right)=\left|\left\langle a_{k}|\Psi_{i}\right\rangle \right|^{2}
\end{equation}
This can be obtained by summing over a complete set of final states, expressing the indifference to the final state. However, we can also arrive at Eq. \ref{rABL} from another direction \cite{Gruss1,Gruss2}. Notice that if the final state is one of the eigenstates, $\Psi_{f}=a_{k}$, then Eq. \ref{ABL} gives probability 1 for measuring $a_{k}$, and probability 0 for measuring any orthogonal state. Consider now an ensemble of systems of which fractions of size $\left|\left\langle a_{k}|\Psi_{i}\right\rangle \right|^{2}$ happen to have the corresponding final states $a_{k}$. The regular probability rule (Eq. \ref{rABL}) of quantum mechanics is then recovered, but now the probabilities are classical probabilities due to ignorance of the specific final states. The same would be the result for a corresponding final state of an auxiliary system, such as a measuring device or environment, correlated with the measured system. This reduction of the ABL rule to the regular probability rule is a clue, showing how a selection of appropriate final states can account for the empirical probabilities of quantum measurements.\\

This possibility of a final state to influence the measurement statistics has motivated the re-formulation of QM as taking into account both initial and final boundary conditions for every system in nature, and ultimately the universe. Within this framework, QM is time-symmetric, in the sense that initial and final states are equally significant in determining the results of a measurement. The Schr\"{o}dinger equation is linear in the time derivative, therefore only one temporal boundary condition may be consistently specified for the wavefunction. If both initial and final boundary conditions exist, we must have two wavefunctions, one for each. The first is the standard wavefunction, or state vector, evolving forward in time from the initial boundary condition. The second is a (possibly) different wavefunction evolving from the final boundary condition backwards in time. Thus, the new formalism is aptly named the TSVF. A measurement, including a post-selection, will later be shown to constitute an effective boundary condition for both wavefunctions. Accordingly, we postulate that the complete description of a closed system is given by the two-state:
\begin{equation}
\label{two-state}
\langle \Phi|~|\Psi\rangle
\end{equation}
where $|\Psi\rangle$ and $\langle \Phi|$ are the forward and backward-evolving states respectively.
These may be combined into an operator form by defining the \textquotedblleft{}two-state density operator\textquotedblright{}:
\begin{equation}
\rho\left(t\right)=\frac{\left|\Psi\left(t\right)\right\rangle \left\langle \Phi\left(t\right)\right|}{\left\langle \Psi\left(t\right)|\Phi\left(t\right)\right\rangle }
\end{equation}
where orthogonal forward and backward-evolving vectors at any time $t$ are forbidden. For a given Hamiltonian $H(t)$, the two-state evolves from time $t_{1}$ to $t_{2}$ according to
\begin{equation}
\rho(t_{2})=U(t_{2},t_{1})\rho(t_{1})U(t_{1},t_{2})
\end{equation}
where $U(t_{2},t_{1})$ is the regular evolution operator:
\begin{equation}\label{TE}
U(t_{2},t_{1})=Texp\left(-i\intop_{t_{1}}^{t_{2}}H\left(\tau\right)d\tau/\hbar\right)
\end{equation}
(T signifies the time ordered expansion). The reduced two-state describing a subsystem is obtained by tracing out the irrelevant degrees of freedom.\\

In standard QM, we may also use an operator form similar to the above, replacing the state vector $\Psi(t)$ with the density matrix:
\begin{equation} \label{rhot}
\rho(t)=\left|\Psi\left(t\right)\right\rangle \left\langle \Psi\left(t\right)\right|
\end{equation}
The density matrix again evolves by Eq. \ref{rhot} and once more the reduced density matrix for a subsystem is obtained by tracing out the irrelevant degrees of freedom. Excluding measurements, the density matrix is a complete description of a system, evolving unitarily from initial to final boundary. Such systems can be thought of as two-time systems having forward and backward-evolving states that are equal at any time, i.e. with a trivial final boundary condition that is just the initial state evolved unitarily from the initial time $t_{i}$ to the final time $t_{f}$,
\begin{equation} \label{psif}
\left|\Psi_{f}\right\rangle =U\left(t_{f},t_{i}\right)\left|\Psi_{i}\right\rangle
\end{equation}
We take Eq. \ref{psif} as a zero-order approximation of the final boundary condition. By considering final boundary conditions deviating from the above, we may introduce a richer state structure into the quantum theory. When would this special final boundary condition show to affect the dynamics, in relation to the trivial boundary condition? It would do so if the reduced two-state describes a subsystem for which the ignored degrees of freedom do not satisfy Eq. \ref{psif}. Then, the reduced two-state should replace the density matrix which is no longer a reliable description of the state of the system. \\

A measurement, as empirically observed, generally yields a new outcome state of the quantum system and the measuring device. This state may be treated as an effective boundary condition for both future, and as indicated by the ABL rule, past events. We suggest it is not the case that a new boundary condition is independently generated at each measurement event by some unclear mechanism. Rather, the final boundary condition of the Universe includes the appropriate final boundary conditions for the measuring devices which would evolve backward in time to select a specific measurement outcome \cite{Gruss1,Gruss2}. In the following sections we shall demonstrate how this boundary condition arises at the time of measurement due to a two-time decoherence effect. Indeed, the ABL rule states that in the pointer basis (determined by decoherence), the outcome of the measurement can only be the single classical state corresponding to the final boundary condition. We thus suggest a particular final boundary condition for the Universe, in which each classical system (measuring device) has, at the time of measurement, a final boundary condition equal to one of its possible classical states (evolved to the final time). \\

A further requirement is that the final state in the pointer basis will induce, backwards in time, an appropriate distribution of outcomes so as to recover the empirical quantum mechanical probabilities for large ensembles, given by the Born rule. The determination of the measurement statistics by the correspondence between the two states may lead one to conclude that within the framework of TSVF, the Born rule is a coincidental state of affairs rather than a law of nature. That is, that the Born rule is a product of an empirically verifiable, yet contingent, relation between the initial and final boundary conditions, one that needs to be postulated for the sake of deducing the rule. This specific relation is contingent in that it leaves open the possibility of a different relation which will lead to a modified version of the Born rule, while the rest of physics remains where it stands. This, however, is inaccurate. It can be shown that this specific law follows, in the infinite $N$ limit, from the compatibility of quantum mechanics with classical-like properties of macroscopic objects \cite{Born,AR}. Under the assumption that for macroscopically large samples, the results of physical experiments are stable against small perturbations, a final state pertaining to the Born rule is the most likely final state, for any ensemble.\\

In addition, the TSVF provides a useful intuition for understanding  several peculiar effects \cite{AAV86,future,unusual}, revealing an unusual but consistent picture of ``weak reality'' \cite{past}. Much experimental work has already verified the predictions of the TSVF by performing weak measurements. At every moment between the pre- and post-selection, weak values, which serve as a generalization of expectation values, can be calculated as follows:
\begin{equation}
\label{weak_values}
\frac{\langle \Phi|A|\Psi\rangle}{\langle \Phi|\Psi\rangle},
\end{equation}
where $A$ is some operator. These values were shown to characterize any weak coupling between two systems when one of them is pre- and post-selected.

\section{The Model}
\label{sec:3}

Let us review our assumptions. The world is described by initial and final boundary conditions. The arrow of time is chosen thermodynamically, that is, by the direction in which entropy increases. We shall assume, according to our current observations, that entropy was low at the beginning of the world and has been increasing since then. In particular, the final state will characterize a highly entangled world with high entropy. Each macroscopic object is comprised of at least $N>>1$ microscopic elements (this number will be specifically defined next), that can be coupled to other, external, microscopic elements. A measurement is a quantum (von Neumann) interaction between the measured system and some microscopic elements within the measuring device, followed by a macroscopic amplification (that is, a record of the result encoded in the state of at least $N$ microscopic particles, or ``environment'' in the language of decoherence). As a result of the measurement a ``collapse'' of the measured microscopic degrees of freedom seems to occur from the experimentalist's point of view. However, those who interpret the result according to the MWI would say that the measured branch was split into several other branches in a unitary way, and hence no collapse has occurred whatsoever. The measurement procedure and apparent collapse will now be analyzed within the TSVF, where the choice of only one branch will be shown to be a consequence of the final boundary condition.

We shall first address the simpler case where only one measurement is performed and the macroscopic world (including measuring devices) stays intact at every time, i.e. does not collapse. Next, we will examine the more general case where more than one measurement is performed. Finally, we will analyze a situation in which the distinction between microscopic and macroscopic world depends only on the number of microscopic elements comprising the objects in question. That is, part of the macroscopic measuring device will be assumed to ``collapse'', but nevertheless, our macroscopic world will be shown to maintain its robustness under time-reversal.\\
Normalizations have been omitted for convenience.

\subsection{A Single ideal measurement}
\label{sec:4}
Let our system be initially described by
\begin{equation}
\label{system_1}
|\Psi(t_0)\rangle = (\alpha |1 \rangle+ \beta |2 \rangle)|READY \rangle |\epsilon_0 \rangle ,
\end{equation}
where $\alpha|1\rangle+\beta|2\rangle$ is the state of a microscopic particle, $|READY\rangle$ is the state of the measuring device, commonly referred to in the literature as ``pointer'' state, and $|\epsilon_0 \rangle$ is the state of the environment. Typically, the device may be macroscopic, but is treated as quantum by the familiar von Neumann scheme. Following this scheme, at time $t=t_1$, we create a coupling between the particle state and the pointer state, establishing a one-to-one correspondence between them. We can generically call the orthogonal pointer states ``I'' and ``II''. The pointer will shift to $|I\rangle$ in case the particle is in $|1\rangle$ and to $|II\rangle$ in case the particle is in $|2\rangle$:
\begin{equation}
\label{coupling_1}
|\Psi(t_1)\rangle = (\alpha |1 \rangle|I \rangle+ \beta |2 \rangle|II \rangle) |\epsilon_0 \rangle.
\end{equation}
Then, in the course of a short time $t_d$, the preferred pointer state is selected and amplified by a multi-particle environment in the process of decoherence. Since the pointer state basis is favored by system-environment interactions, it is not prone to further entanglement and decoherence. Therefore, it enables us to read off the result of the measurement from the environment in which it is encoded in a unitary fashion. The reading of a specific result does not correspond to just one specific state of the apparatus/environment but rather to a subset of states taken from a very large state-space, where distinct readings correspond to orthogonal states. Physically, these may be spatially separated blotches on a photo-detector, or concentration of molecules in a corner of a chamber. We represent these distinct environmental subsets as $\epsilon_1$ and $\epsilon_2$, and the dynamical process is thus
\begin{equation}
\label{decoherence_1}
|\Psi(t_1+t_d)\rangle = \alpha |1 \rangle|I \rangle|\epsilon_1 \rangle+ \beta |2 \rangle|II \rangle|\epsilon_2 \rangle ,
\end{equation}
This is a macroscopic amplification of the microscopic measurement, which results in what we call ``measurement'' of the particle. After this point, the particle may continue to interact with other objects (microscopic or macroscopic).\\
Now comes the crucial part. Let the backward-evolving state at $t=t_f$ contain only a single term out of the preferred pointer basis
\begin{equation}
\label{final_1}
\left\langle \Phi\left(t_f\right)\right|= \langle\phi| \langle I| \langle \epsilon_1|,
\end{equation}
where $\langle\phi|$ is a final state of the microscopic particle. Within the TSVF, our system will be described by the two-state:
\begin{equation}
\label{TSVF_1}
\langle\phi| \langle I| \langle \epsilon_1|~(\alpha |1 \rangle|I \rangle|\epsilon_1 \rangle+ \beta |2 \rangle|II \rangle|\epsilon_2 \rangle),
\end{equation}
for $t_1+t_d<t<t_2$. This is essentially a future choice of $|I\rangle$, which may serve as a reason for the initial outcome represented by the microscopic state $|1\rangle$. The approximate orthogonality of $|\epsilon_1\rangle$ and $|\epsilon_2\rangle$ assures that after reducing the density matrix to include only the observable degrees of freedom, within the interval $t_1+t_d<t<t_2$, only the first term in Eq. \ref{TSVF_1} will contribute, accounting for the macroscopic result we witness. In the most general case, the backward environment-pointer state need not be exactly identical to the corresponding term in the forward-state, as long as the measure of its projection on it is exponentially (in the number of particles) larger than the measure of its projection on the non-corresponding term(s) of the forward state.\\

Regarding weak values (Eq. \ref{weak_values}), if any were measured during the intermediate times, they would have been determined by the specific selection made by the final state. Moreover, any interaction with this pre- and post-selected ensemble would reflect this final state in the form of the ``weak potential'' \cite{unusual}.\\

The effective boundary condition for the past of the backward-evolving state determines the observed measurement outcome by a backward decoherence process. That is, just the same as the backward-state sets the boundary for the future of the forward-evolving state, the forward-evolving state sets the boundary for the past of the backward-state. Together with the regular decoherence, this amounts to a symmetric two-time decoherence process \cite{Gruss1,Gruss2}, allowing for a generalization to multiple-time measurements. This subject is formalized in the next subsection, where we present a detailed description of two consecutive measurements. \\

The important conclusion we should bear in mind is related to the time-reversed process. Starting from the state $|\Psi(t_f)\rangle$ as described by Eq. \ref{final_1} and going backwards in time, we are able to reconstruct the pointer reading $|I\rangle$ although the measured microscopic particle has changed its state. This relates to the concept of ``macroscopic robustness under time-reversal'' we elaborate on in Subsection \ref{sec:6}. \\

We note that in cases where the free Hamiltonian is non-zero, we will have to apply the forward time-evolution operator of Eq. \ref{TE} on the final boundary condition, which would then cancel upon backward time evolution to the present state. This clearly does not change the results, and therefore we preferred to discuss a zero Hamiltonian.\\

\subsection{Multiple measurements within the two-time decoherence scheme}
\label{sec:5}
We shall demonstrate how a measurement effectively sets appropriate boundary conditions for past and future measurements, and how a definite macro-history for subsequent measurements can be drawn out by suitably chosen boundary conditions. For deductive purposes, we will lay out the scheme for performing two consecutive measurements rather than one. This model can be conveniently generalized toward $n$ consecutive measurements (a limitation on this figure will be discussed later), as well as for continuous observables and mixed states.\\

The possibility to draw such a scheme is important also since two sequential measurements performed in different bases seem to deny time reversibility in a single branch of the wavefunction. It is clear that some of the information gathered in the first measurement will be lost after performing the second. However, we shall demonstrate how the final boundary condition correctly restores the two consecutive readings, and thus grants our macroscopic experience robustness.\\

In order to emphasize the generality of our method, we will write the calculation this time using density matrices and in decoherence notation. \\

Consider an experiment performed on a spin-half particle, in order to measure its spin component along some axis. Let the initial state of the particle be $a\left|\uparrow\right\rangle _{x}+b\left|\downarrow\right\rangle _{x}$,
and denote, as before, the initial state of the pointer ``READY'' or ``R''. There are 2 measuring devices since we are about to perform 2 consecutive measurements (we can perform the two measurements also with a single device,
and include an initialization process between the measurements). The states of the environment are labeled in accordance with the pointer readings they indicate. For example, the sub-index \\
$x(R)y(U)$ means ``READY'' in the $x$ axis and ``UP'' in the $y$ axis. We assume that states indicating different readings are orthogonal to each other.\\

The state of the composite system and environment at the initial time $t_{0}$ is
\begin{equation}
\left|\Psi\left(t_{0}\right)\right\rangle =\left(a\left|\uparrow\right\rangle _{x}+b\left|\downarrow\right\rangle _{x}\right)\left|R\right\rangle _{x}\left|R\right\rangle _{y}\left|\epsilon_{x(R)y(R)}\right\rangle
\end{equation}
We set up a device to measure the spin along the $x$ axis. The coupling interaction lasts $t_{I}$, and by the end of it, the system is unitarily transformed to the state:
\begin{equation}
\left|\Psi\left(t_{1}\right)\right\rangle =\left(a\left|\uparrow\right\rangle _{x}\left|U\right\rangle _{x}+b\left|\downarrow\right\rangle _{x}\left|D\right\rangle _{x}\right)\left|R\right\rangle _{y}\left|\epsilon_{x(R)y(R)}\right\rangle,
\end{equation}
where $\left|U\right\rangle _{x}$ and $\left|D\right\rangle _{x}$ are orthogonal. After decoherence takes place, the pointer is also entangled with some of the environmental degrees of freedom, resulting in
\begin{equation}
\begin{array} {lcl}
|\Psi(t_{1}+t_{d})\rangle=(a|\uparrow\rangle _{x}|U\rangle _{x}|\epsilon_{x(U)y(R)}\rangle \\
+b|\downarrow\rangle _{x}|D\rangle _{x}|\epsilon_{x(D)y(R)}\rangle)|R\rangle _{y}
\end{array}
\end{equation}
Decoherence is assumed to cause these states to remain classical up to some far \textquotedblleft{}final time\textquotedblright{}. For the time being, assume that after the measurement interaction is over, the measuring device is left idle and its state remains unchanged. Let us suppose for the moment that the backward-evolving state singles out the ``UP'' pointer state at  $t_{1}+t_{d}$. However, since globally no collapse has taken place, we allow the ``DOWN'' term to persist. We will immediately show that it has no physical manifestation.\\

Now we set up the second device to measure the spin along the $y$ axis. The interaction changes it to
\begin{equation}
\begin{array}  {lcl}
\left|\Psi\left(t_{2}\right)\right\rangle = \frac{1}{\sqrt{2}}\biggl(a\biggl(\left|\uparrow\right\rangle _{y}\left|U\right\rangle _{x}\left|\epsilon_{x(U)y(R)}\right\rangle \left|U\right\rangle _{y} \\
+ \left|\downarrow\right\rangle _{y}\left|U\right\rangle _{x}\left|\epsilon_{x(U)y(R)}\right\rangle \left|D\right\rangle _{y}\biggr) \\
+ b\biggl(\left|\uparrow\right\rangle _{y}\left|D\right\rangle _{x}\left|\epsilon_{x(D)y(R)}\right\rangle \left|U\right\rangle _{y} \\
+\left|\downarrow\right\rangle _{y}\left|D\right\rangle _{x}\left|\epsilon_{x(D)y(R)}\right\rangle \left|D\right\rangle _{y}\biggr)\biggr)
\end{array}
\end{equation}
and after decoherence:
\begin{equation}
\begin{array}  {lcl}
\left|\Psi\left(t_{2}+t_{d}\right)\right\rangle = \frac{1}{\sqrt{2}}\biggl(a\biggl(\left|\uparrow\right\rangle _{y}\left|U\right\rangle _{x}\left|U\right\rangle _{y}\left|\epsilon_{x(U)y(U)}\right\rangle \\
+\left|\downarrow\right\rangle _{y}\left|U\right\rangle _{x}\left|D\right\rangle _{y}\left|\epsilon_{x(U)y(D)}\right\rangle \biggr) \\
+ b\biggl(\left|\uparrow\right\rangle _{y}\left|D\right\rangle _{x}\left|U\right\rangle _{y}\left|\epsilon_{x(D)y(U)}\right\rangle \\
+\left|\downarrow\right\rangle _{y}\left|D\right\rangle _{x}\left|D\right\rangle _{y}\left|\epsilon_{x(D)y(D)}\right\rangle \biggr)\biggr)
\end{array}
\end{equation}
Let us consider at that final time a backward-evolving state:
\begin{equation}
\begin{array}  {lcl}
\left\langle \Phi\left(t_{2}+t_{d}\right)\right|=\left\langle \varphi\right|\left\langle U\right|_{x}\left\langle U\right|_{y}\left\langle \epsilon_{x(U)y(U)}\right|
\end{array}
\end{equation}
At $t_{2}+t_{d}$, the complete description of the composite system is given by
\begin{equation}
\begin{array}  {lcl}
\rho\left(t_{2}+t_{d}\right)= \frac{1}{\sqrt{2}}\biggl(a\biggl(\left|\uparrow\right\rangle _{y}\left|U\right\rangle _{x}\left|U\right\rangle _{y}\left|\epsilon_{x(U)y(U)}\right\rangle \\
+\left|\downarrow\right\rangle _{y}\left|U\right\rangle _{x}\left|D\right\rangle _{y}\left|\epsilon_{x(U)y(D)}\right\rangle \biggr) \\
+b\biggl(\left|\uparrow\right\rangle _{y}\left|D\right\rangle _{x}\left|U\right\rangle _{y}\left|\epsilon_{x(D)y(U)}\right\rangle \\
+\left|\downarrow\right\rangle _{y}\left|D\right\rangle _{x}\left|D\right\rangle _{y}\left|\epsilon_{x(D)y(D)}\right\rangle \biggr)\biggr)
\left\langle \varphi\right|\left\langle U\right|_{x}\left\langle U\right|_{y}\left\langle \epsilon_{x(U)y(U)}\right |
\end{array}
\end{equation}
And the reduced density matrix:
\begin{equation}
\begin{array} {lcl}
\rho_{reduced}(t_{2}+t_{d})=Tr_{\epsilon}\rho\simeq
|\uparrow\rangle _{y}|U\rangle _{x}|U\rangle_{y}\langle \varphi|\langle U|_{x}\langle U|_{y}
\end{array}
\end{equation}
The environment singles out the ``UP'' pointer state from this time onwards, and sets the forward-evolving spin at $|\uparrow\rangle_y$, as expected from a $y$ ``UP'' measurement. Due to the reduction, the other terms of the forward and backward-evolving states are not experimentally observed, causing the evolution to appear non-unitary, hence creating an effective ``collapse''. \\

What about the intermediate time between the 2 measurements? If the $y$ spin coupling interaction is applied to the backward-evolving state at time $t_{2}$, we arrive at
\begin{equation}
\left\langle \Phi\left(t_{2}-t_{I}\right)\right|=\left(c\left\langle \uparrow\right|_{y}\left\langle R\right|_{y}+d\left\langle\downarrow\right|_y\left\langle O\right|_{y}\right)\left\langle U\right|_{x}\left\langle \epsilon_{x(U)y(U)}\right|
\end{equation}
for some parameters $c$ and $d$, where the time-reversed interaction between the measuring device and the particle, causes a device in the final state ``UP'' to evolve backwards into the state ``READY'', if the particle is in the state  $\uparrow$, and into the orthogonal state ``ORTHO'', if the particle is in the state $\downarrow$. Again we assume that an environment-induced decoherence and selection of pointer states takes place (here backwards in time, but the microscopic physics is time-symmetric), singling out ``READY'' and ``ORTHO'' as the preferred basis of pointer states for the backward-evolving vector, resulting in
\begin{equation}
\begin{array}  {lcl}
\langle \Phi(t_{1}+t_{d}<t<t_{2}-(t_{I}+t_{d}))|=\\
\biggl(c\langle \uparrow|_{y}\langle R|_{y}\langle \epsilon_{x(U)y(R)}|+d\langle \downarrow|_y\langle O|_{y}\langle \epsilon_{x(U)y(O)}|\biggr)\langle U|_{x}
\end{array}
\end{equation}
The composite system at the intermediate time is therefore:
\begin{equation}
\begin{array}  {lcl}
\rho(t_{1}+t_{d}<t<t_{2}-(t_{I}+t_{d}))=\\
\biggl(a|\uparrow\rangle _{x}|U\rangle _{x}|R\rangle _{y}|\epsilon_{x(U)y(R)}\rangle \\
+b|\downarrow\rangle _{x}|D\rangle _{x}|R\rangle _{y}|\epsilon_{x(D)y(R)}\rangle\biggr)\\
\biggl(c\langle\uparrow|_{y}\langle U|_{x}\langle R|_{y}\langle\epsilon_{x(U)y(R)}|\\
+d\langle\downarrow|_{y}\langle U|_{x}\langle O|_{y}\langle\epsilon_{x(U)y(O)}|\biggr)
\end{array}
\end{equation}
And the reduced density matrix is given by
\begin{equation}
\begin{array}  {lcl}
\rho_{reduced}(t_{1}+t_{d}<t<t_{2}-(t_{I}+t_{d}))=Tr_{\epsilon}\rho\simeq\\
|\uparrow\rangle _{x}|U\rangle _{x}|R\rangle _{y}\langle \uparrow|_{y}\langle U|_{x}\langle R|_{y}
\end{array}
\end{equation}
In this time interval, the effective reduction has singled out the pointer state ``UP'' in $x$, so that measurements on the environment will consistently give ``UP''. Here we can see how the environment, mediated by the pointer, is responsible for transmission of the particle spin state backwards in time (through backward decoherence), establishing a boundary condition for any past measurement. Information for the reduction of the backward-evolving state is carried by the measuring device backward-evolving state, and the rest of the environment in which it is encoded. The forward-evolving state of the particle before the time of the measurement is of course not affected by the final boundary condition. Proceeding in the same manner, this scheme can easily be shown to establish an effective backward-evolving state $\left\langle \uparrow\right|_{x}$, setting a final boundary condition for any measurement performed on the particle at $t<t_{1}-\left(t_{I}+t_{d}\right)$.\\

To conclude, a two-time decoherence process is responsible for setting both forward and backward boundaries of the spin state to match the result of a given measurement, allowing multiple-time measurements to be accounted for by our model.

\subsection{Macroscopic world also decays}
\label{sec:6}

We would like to address the issue of time reversibility of the dynamical equations governing the macrostates. While this property is most naturally present in the MWI, as long as the macroscopic objects stay intact, it may also exist in a single branch, even if its history includes non-unitary events. This is due to the fact that the result of a measurement performed on a microscopic state is stably stored within the macroscopic objects, as we have seen in the last chapter, and can theoretically be extracted. Therefore, while the measured microstate may change non-unitarily from our local perspective, our measurement reading may not. This possibility is what we will refer to as ``classical robustness under time-reversal''. As well as being a landmark of classical physics, time reversibility is vital in order to draw valid conclusions about the early Universe from our current observations.\\

Let the system not be completely isolated, and allow external quantum disturbances which interfere with the evolution in an indeterministic and thus irreversible way (from the single-branch perspective). It is generally accepted that the pointer states selected by the environment are immune to decoherence, and are naturally stable \cite{Zurek,Deco}. Problems start when the (macroscopic) measurement devices begin to disintegrate to their microscopic constituents, which may couple to other macroscopic objects and effectively ``collapse". These collapses seem fatal from the time-reversal perspective, as time-reversed evolution would obviously give rise to initial states very different from the original one. To tackle this, we demand that subtle environmental interactions, mildly altering the macrostate, will not stray too far from the subset of states indicating the perceived measurement result, compared to the orthogonal result.\\

Considering the free evolution of the measuring device and applying it backwards from the final and slightly altered state, the state at the time of measurement will still project heavily onto the same sub-space, indicating the same reading. This may be regarded as macroscopic physics having time-symmetric dynamics. While it might be the case that we do not reconstruct the starting microscopic configuration, being macroscopic objects, this should not upset us as long as our experience remains the same, that is, as long as macroscopic readings, depending on the macrostate of their $N$ micro-particles, do not change when backward evolution is applied. This will be shown to be the case when several assumptions are made regarding the macroscopic objects and the rate of collapse.\\

To derive the limit between microscopic and macroscopic regimes we will assume now that the amplification mechanism consists of at least $N>>1$ particles belonging to the environment or measuring device, from which only $n<<N$ particles may later be measured and collapsed without rendering the dynamics irreversible. by ``measured" we do not necessarily mean that an observer entangled them with a device designated for measurement. Rather, we mean that they may get entangled with some other degree of freedom and decohere. We believe that it is reasonable to assume that $n<<N$ always, because measuring $N$ (which is typically, $10^{20}$) particles and recording their state is practically impossible. Eqs. \ref{system_1}-\ref{decoherence_1} still have the same form, but the measurement of the environment at some $t=t_2$ leads to
\begin{equation}
\label{particle_2}
|\epsilon_i(N)\rangle~~\rightarrow~~\prod_{j=1}^{N-n}|C_i^{(j)}\rangle|\epsilon_i(N-n)\rangle
\end{equation}
for $i=1,2$ representing encodings of two orthogonal microstates. The environment states are $N$-particles states at first, and later contain only $N-n$ particles, while their other $n$ components ``collapse'', for simplicity of calculation, to the product state $\prod_{j=1}^{n}|C_i^{(j)}\rangle$.
The trivial point, although essential, is that
\begin{equation}
\langle C_1^{(j)}|\epsilon_1^{(j)}\rangle =\gamma_1^{(j)} \neq 0,
\end{equation}
for every $j=1,2,...,n$, where $\epsilon_1^{(j)}$ is the j-th environment state before the collapse, i.e. collapse can never reach an orthogonal state. For later purposes let us also assume
\begin{equation}
\langle C_2^{(j)}|\epsilon_1^{(j)}\rangle =\gamma_2^{(j)} \neq 0
\end{equation}
It is not necessarily different from $0$, but as will be demonstrated below, this is the more interesting case. We would like to show that the final boundary state of Eq. \ref{final_1} still has much higher probability to meet $|\epsilon_1\rangle$ than $|\epsilon_2\rangle$, and hence the pointer reading is determined again by the specific boundary condition, despite the collapse of the classical apparatus. Indeed, under the assumption of ending the evolution in the following final boundary condition:
\begin{equation}
\label{final_2}
\left\langle \Phi\left(t_f\right)\right|= \langle\phi| \langle I| \langle \epsilon_1|,
\end{equation}
we can define the ``robustness ratio'' as a ratio of probabilities: The probability to reach backwards in time the ``right'' state $|I\rangle$ divided by the probability to reach the (``wrong'') $|II\rangle$ state. This ratio ranges from zero to infinity suggesting low (values smaller than $1$) or high agreement (values grater than $1$) with our classical experience in retrospect. In our case it is
\begin{equation}
\begin{array} {lcl}
\label{RR}
\frac{Pr(Right)}{Pr(Wrong)}=\frac{\prod_{j=1}^n\gamma_1^{(j)}}{|\langle \epsilon_1(N-n)| \epsilon_2(N-n)\rangle|^2\prod_{j=1}^n\gamma_2^{(j)}}\simeq \\ \\
\simeq |\langle \epsilon_1(N-n)| \epsilon_2(N-n)\rangle|^{-2}
\end{array}
\end{equation}
Hence for a sufficiently large ratio of $N/n$ ``classical robustness'' is attained - the result of Eq. \ref{RR} is exponentially high (and even diverging if we allow some $\gamma_2^{(j)}$ to be zero). The significance of the result is the following: even though from the perspective of the single branch a non-unitary evolution has occurred, there exists a final boundary state which can reproduce with high certainty the desired macroscopic reality when evolved backwards in time. This ``robustness ratio'' can be used also for the definition of macroscopic objects, i.e. defining the border between classical and quantum regimes.\\

The MWI was invoked in order to eliminate the apparent collapse from the unitary description of QM. Within the MWI, the dynamics of the universe is both symmetric and unitary. We have now shown that these valuable properties can be attained even at the level of a single branch, that is, without the need of many worlds, when discussing macroscopic objects under suitable boundary conditions. Despite the seemingly non-unitary evolution of microscopic particles at the single branch, macroscopic events can be restored from the final boundary condition backwards in time due to the encoding of their many degrees of freedom in the final state.


%
%

\section{Discussion}
\label{Disc}
We have seen that at the cost of having to introduce a final boundary condition, we are able to reclaim determinism and ensure macroscopic time-reversal. It was already assumed, that subsequent to the measurement interaction, decoherence causes an effectively irreversible branching of the superposition into isolated terms. Therefore, no inconsistencies can arise from the existence of a special final boundary condition of the form described before, which simply causes the selection of a single specific branch from the many worlds picture. In this view, the measurement process does not increase the measure of irreversibility beyond that of regular thermodynamics. Additionally, accounting for the apparent collapse, TSVF does not suggest a microscopic quantum mechanical arrow of time. It does however assume asymmetric initial and final boundary conditions.\\

It must be emphasized that the final boundary condition for any measurement, which we have taken to be specific in the examples, is generally unknown prior to the completion of that measurement. In fact, we can never simultaneously determine both forward and backward states of our target microscopic degree of freedom, which is why we cannot gain more complete knowledge of future measurement outcomes than what is offered by the standard formalism. Note that the measurements set a boundary for the forward-evolving state of the microscopic degree of freedom at $t>t_{meas}$, and for the backward-evolving state in $t<t_{meas}$ (see Subsection \ref{sec:5}), but not both at the same time interval.\\

Classically, an {\it a priori} knowledge of the future is of course an acausal state of affairs. The following example shows that it is also a problem due to quantum mechanics itself. Assume there are two entangled spin-half particles located at two far away locations, in the initial state
\begin{equation}
|\Psi_i\rangle=\frac{1}{\sqrt{2}}(|\uparrow_A\rangle|\uparrow_B\rangle + |\downarrow_A\rangle |\downarrow_B\rangle),
\end{equation}
where, ``$A$'' denotes the particle at Alice's location, and ``$B$'' the particle at Bob's location. Assume that Bob knows the final state to be
\begin{equation}
|\Psi_f\rangle=\frac{1}{\sqrt{2}}(|\uparrow_A\rangle|\uparrow_B\rangle + |\uparrow_A\rangle |\downarrow_B\rangle)
\end{equation}
Now, Alice may or may not perform a unitary rotation on her particle, say $\sigma_x$, under which
 \begin{equation}
 \begin{array} {lcl}
 | \uparrow_A \rangle \longrightarrow | \downarrow_A \rangle \\
 |\downarrow_A\rangle \longrightarrow |\uparrow_A\rangle
\end{array}
\end{equation}
Leaving the initial composite state as it was, or transforming it into the state
\begin{equation}
|\Psi_t\rangle=\frac{1}{\sqrt{2}}(|\downarrow_A\rangle|\uparrow_B\rangle + |\uparrow_A\rangle |\downarrow_B\rangle)
\end{equation}
Bob, measuring the spin of his particle, obtains $|\downarrow_B\rangle$ or $|\uparrow_B\rangle$, according to the action or non-action of Alice. In this manner, Alice may transmit signals to Bob at an instant. A procedure like this would be possible for many arbitrary choices of initial and final states. Only when identical measurements are performed in sequence, can a final state (or a measurement outcome) be predicted with certainty in advance. Therefore, as in hidden variables theories, the parameters determining the measurement outcome, in this case the final state, must remain inaccessible.\\

Still, as in those theories, the evolution may be considered deterministic (though unpredictable). As mentioned before, determinism is valid if considered in a broader two-time context, where the evolution is determined not only by an initial boundary condition, but also by a final boundary condition. The latter dynamically determines the measurement outcomes, such that in all intermediate times the physical states are determined by two unitary time-evolutions. Given the boundary conditions and a Hamiltonian, one may reconstruct the whole evolution history, no random dice need be tossed. Alternatively, the state of the system at any given time is completely determined by its two-state at any single time in the course of history.\\

Another property we wish to address is that of ``realism'' or ``objectivity''. These refer to the classical concept that the existence of physical properties is independent of observations of these properties. EPR \cite{EPR} define realism by the following counterfactual:\\

\textit{``If, without in any way disturbing a system, we can predict with certainty (i.e., with probability equal to unity) the value of a physical quantity, then there exists an element of physical reality corresponding to this physical quantity''}.\\

This would require the existence of some additional (possibly ``hidden'') variables, which determine the outcomes of the measurements. As mentioned before, hidden variable theories are inherently nonlocal \cite{Bell}, and the possibility of local realism is excluded.\\

This line of reasoning requires the validity of ``counterfactual definiteness''. The consequence of which is that it is meaningful to ask hypothetic ``what-if'' questions. Otherwise, the EPR definition of realism is irrelevant. Such is the case with the many worlds interpretation, where each measurement yields all possible outcomes. This is also the case with the suggested interpretation, where it is assumed that the final boundary condition determines the preferred basis for the measurement device and chooses the outcomes backwards in time. Therefore, the backward-evolving state may be viewed as constituting a hidden variable, which answers only the question that it is being asked. From our perspective as remembering only the past (an issue which remains to be explained), .\\

In addition, we would like to address some potential criticism (See \cite{TMWI} for example):\\

1. ``The final state is very unique and is thus unlikely to occur spontaneously, making this construction rather artificial.''.\\

According to the TSVF, any post-selected state which is not orthogonal to the pre-selected state is permissible. However, in our model we have discerned a special boundary condition which accounts for the experimental result, i.e. for the single outcomes which actually occurred in measurements that were actually performed, as well as for the Born rule statistics. This choice is justified on several grounds:\\
1. It unites the two dynamical processes of textbook quantum mechanics - the SE and collapse - under one heading. In doing so, it renders QM deterministic and local on a global level, and above all, rids of the ambiguity involved in the ``yes SE but no SE'' approach. \\
2. It allows for robustness under time-reversal of macroscopically large systems.\\
3. It is a natural framework to understand weak values and weak reality.\\
4. As explained, it is not the case that we could have chosen any sort of boundary condition and still maintain classicality on the macro-level. Only states pertaining to the Born rule allow for that. So empirical observations other than the Born rule by itself (e.g. stability under perturbations) can be seen as supporting evidence for a backwards-evolving state with these properties, if any at all.\\
We find these reason enough to postulate such a boundary condition. Moreover, the initial state of our universe can also be regarded as ``unique'', and therefore, we would like to perceive these two boundary conditions as reasonable, constructive and even necessary for explaining our current observations, rather than artificial. It should also be stressed that in spite of this ``uniqueness'', the final state has high thermodynamical entropy and also high entanglement entropy, since it encodes all measurement outcomes of microscopic objects.\\

2.``After a long time, $n$ becomes comparable to $N$, thus failing the above reasoning''. The case of $n$ consecutive measurements is discussed in Subsection \ref{sec:5}. This case follows the same logic of only one measurement and is fully consistent with classical robustness. In addition, we claim that $n$ cannot grow to be $N$, i.e., there is always a ``macroscopic core'' to every macroscopic object which contained initially $N$ or more particles. It can be shown that $\frac{dn}{dt} \le 0$ and also that $\frac{dn}{dt} \rightarrow 0$ for long enough times, assuming for example an exponential decay of the form:
\begin{equation}
N(t)=N(0)exp(-t/T),
\end{equation}
where $T$ is some constant determining the life time of macroscopic objects. Also, on a cosmological scale (inflation of the universe) it can be shown that after long time, measurements become less and less frequent (macroscopic objects which can perform measurements are simply no longer available). That means there is more than one mechanism responsible for a finite number (and even smaller than $N$) of collapses at any finite or infinite time of our system's evolution.\\

Finally, some philosophical reflections. In the context of time-symmetric formulations of quantum mechanics, it is argued for many years now, that God plays dice in order to save free will \cite{Gruss1,Gruss2,AR}. This claim was now analyzed from another point of view, and the heavenly dice is shown to account for the Born rule in the microscopic world even though it is restricted macroscopically by the final boundary condition.\\

It is common philosophical practice to point out the tension between the concepts of free will and determinism. One of the virtues of the TSVF is that it gives rise to a new refined version of determinism, which sheds fresh light on the relation between these seemingly conflicting ideas. Within this framework, while both backward and forward states evolve deterministically, they have limited physical significance on their own - the physical reality is the product of the causal chains extending in both temporal directions. The past does not determine the future, yet the future is set, and only together do they form the present. But the existence of a future boundary condition, and its deterministic effect, do not deny our freedom of choice. It is allowed due to the inaccessibility of the data (which was a requirement of causality, as shown). Examining the concept of free will from a physical point of view, we find it must contain at least partial freedom from past causal constraints, and such freedom is duly manifested in the TSVF, where a juxtaposition of freedom and determinacy is epitomized \cite{Gruss1,Gruss2,AR}. Being macroscopical objects composed of many microscopic particles, we enjoy benefits from both worlds: freedom from the quantum domain and determinism from the classical domain, ensured by robustness.\\

All in all, we find this resolution very appealing because it is parsimonious (does not involve a multitude of worlds), deterministic (in a two-time sense) and fits our observations on either microscopic or macroscopic scales.
The complementary view we plan to discuss in future works, is depicted by the notion of ``Each instant of time a new universe'' \cite{each}.

\begin{acknowledgements}
This work has been supported in part by the Israel Science Foundation Grant No. 1125/10 and by the ICORE Excellence Center ``Circle of Light''. We would like to thank Avshalom C. Elitzur, Boaz Tamir, Lior Deutsch and Omer Goldman for helpful discussions. We would also like to thank an anonymous referee for many helpful remarks.
\end{acknowledgements}




\end{document}